\def\approxlt{\lower.2em\hbox{$\buildrel < \over \sim$}}
\def \kms{\hbox{km$\,$s$^{-1}$}}
\def \ls{\hbox{L$_{\odot}$}}
\def \Lsun{\hbox{L$_{\odot}$}}
\def \Msun{\hbox{M$_{\odot}$}}

\def \etal{{et~al. }}

\def\date{\number\day\space \ifcase\month\or
        January\or February\or March\or April\or May\or June\or 
        July\or August\or September\or October\or November\or December\fi
        \space\number \year}
\documentstyle[12pt,aasms4]{article}
\begin{document}
\title{MOLECULAR GAS  IN THE SPECTACULAR RING GALAXY NGC~1144  }
\author{Yu Gao\footnote{current address: 
Laboratory for Astronomical Imaging, Department of Astronomy,  
University of Illinois, 1002 West Green Street, Urbana, IL 61801}
and P.M. Solomon}\affil{Astronomy Program, State University of New
York, Stony Brook, NY 11794-2100}
\author{D. Downes} \affil{Institut de Radio Astronomie Millim\'etrique, 
    38406 St. Martin d'H\`eres, France}
\and\author{S.J.E. Radford}
\affil{National Radio Astronomy Observatory, Tucson, AZ 85721-0665}

\received{September 25, 1996}
\begin{abstract}

We have detected extremely wide (1100 \kms)
CO(1--0) emission from NGC 1144, an interacting, luminous
infrared galaxy that is the dominant component of the
Arp 118 system.  The observations show that
NGC 1144 is one of the most CO luminous galaxies in the local
universe, with a CO luminosity twice that of Arp~220.
Maps with the IRAM interferometer show that 
the CO is not in or very near the Seyfert~2 nucleus, but in the 20\,kpc diameter 
ring that extends 
halfway between NGC 1144 and the elliptical galaxy NGC 1143. 
The greatest gas concentration, with 40\% of the CO luminosity,
is in the southern part of the ring,  in NGC 1144.
Another 15\% of the CO luminosity comes from the dominant 10\,$\mu$m source,
a giant extranuclear HII  region.
 The ring of molecular gas, the off-center nucleus, the
ring extending halfway to the intruder, and the velocity of the intruder 
nearly equal to the escape velocity all show that Arp~118 is a ring galaxy
produced by a  collision of a massive spiral with an elliptical. 
The most spectacular property is the velocity range, which in
 Arp~118 is 2 to 3 times higher than in a typical ring galaxy.  Arp~118 is a rare example
 of a very luminous extended starburst with a scale of about 5--10 kpc, and a 
luminosity of 3 $\times \,10^{11}$\ls .

\end{abstract}


\section{INTRODUCTION}
Most luminous infrared galaxies are merging or interacting with
 strong nuclear concentrations of molecular gas. 
There is, however, another class of interacting galaxies, ring galaxies  
with non-nuclear but coherent starbursts.  The Arp 118 system has long been 
recognized as a member of this class (e.g., Freeman \& de Vaucouleurs 1974).  
Its IR luminosity  
\footnote{ for $H_0=75$ \kms \,Mpc$^{-1}$, the distance to Arp~118 is 118\,Mpc.}
,
$L_{{\rm IR}} = 2.5 \times 10^{11} \ {\rm L}_{\odot}$,
is one of the highest, about 10 times that of other ring galaxies
(Appleton \& Struck-Marcell 1987).  
The Arp 118 pair consists of a disk galaxy, NGC 1144, that has
interacted with NGC 1143, an elliptical galaxy $40''$  
northwest (Fig.~1). NGC 1144 has a strongly distorted disk, two 
bright optical sources in its central regions, a stellar bridge
seen at 2\,$\mu$m between the two galaxies (Joy \& Ghigo 1988 [JG88]),
and a ring of HII regions extending halfway between NGC 1144 and NGC 1143  
(Hippelein 1989 [H89]). 

Arp 118's morphology and velocity spread indicate a strong interaction. 
Although the systemic velocities of the nuclei of 
NGC 1144 and NGC 1143 differ by only 
300\,\kms, the internal velocity spread in NGC 1144 is much higher. 
Observations of H$\alpha + [{\rm NII}]$ and other lines show velocity
shifts over 1100\,\kms\ (H89). The strongest starburst is a 
giant HII region complex $8''$ west of the nucleus. This 
giant HII region dominates the
galaxy's 10\,$\mu$m and H$\alpha$ luminosities even though
NGC~1144 has a Seyfert 2 nucleus (H89; Osterbrock \& Martel 1993).
There are several non-thermal radio sources in 
NGC 1144, including a very compact radio source at the nucleus, and a strong
extended region 5$''$ to the northeast (JG88; Condon \etal 1990; Fig.~1). 

In contrast to the higher-energy activity traced by the
relativistic particles and the ionized gas, 
the far IR colors indicate cold dust at 34 K. To see whether the cold,
massive, molecular gas showed the same extreme kinematics as the
ionized gas, we made molecular line observations with the IRAM 30\,m
telescope and the IRAM interferometer.

\section{OBSERVATIONS AND  DISTRIBUTION OF THE MOLECULAR GAS}
CO observations were made  with the IRAM 30m 
telescope on Pico Veleta near Granada, Spain. We used 
$512 \times 1$ MHz filterbanks and SIS receivers with 
SSB system temperatures of 250 and 
450\,K ($T_{\rm a}^*$), respectively, for 
CO(1--0) and CO(2--1). Both lines were observed simultaneously. 
 Pointing was checked by observing
nearby quasars every two hours, with errors $\sim 3''$.
We also looked for 
HCN(1--0) and CS(3--2), with system temperatures of 270 
and 300\,K, respectively.  The beamwidths were $23''$ and 
$28''$ for CO(1--0) and HCN(1--0). 
CO was observed at three positions in Arp 118
(Fig.~1) and HCN was observed only 
toward NGC 1144. Integration times were 30 to 60 min  
for each position in CO and 4 hours in HCN.
The 500\,MHz spectrometer
bandwidth barely covered the very wide lines in CO(1--0), and
CO(2--1) was used only to confirm detections.

CO(1--0) was also observed with the four 15m antennas of the IRAM  
Interferometer on Plateau de Bure, France.  
Three configurations gave a synthesized beam of 
$5.3'' \times 2.5''$  at position angle (p.a.) $20^{\circ}$. 
The SIS receivers had SSB system temperatures 
of 300\,K. The spectral 
coverage was 480\,MHz (1285\,\kms) with a resolution of 2.5\,MHz (6.7 
\kms). Since the interferometer's field of view is $45''$ (FWHM)
at 3\,mm, we mapped Arp 118 with two overlapping fields. 
The first field covered NGC 1144 and the second 
covered the northwest part of the ring and NGC 1143.

The single-dish CO line profile in NGC 1144 is remarkably strong and wide, 
$\sim$ 1100 \kms \ to zero intensity (Fig.~1). 
This is the widest CO line ever detected.  
CO was also detected in the northwest part of the ring,
at $17''$ (10 kpc) from the main disk of NGC 1144.  At this position,
the single-dish CO spectrum  peaks at $\sim 8250$ \kms\  (Fig.~1).
 No CO was detected in the elliptical NGC 1143. 
Weak HCN(1--0) (not shown here) was marginally detected in NGC 1144 at 
9100 \kms
    The implied HCN-to-CO intensity ratio in NGC 1144 
is comparable to that of normal spiral galaxies (Solomon \etal 1992).
CS(3--2) was not detected.

The CO emission near 9100 \kms \ in the single-dish spectrum of
NGC~1144 (Fig.~1) 
corresponds to the strong CO sources 6 and 7 detected by the interferometer 
in the south of the ring (Figs.~2 and 3). 
The emission near 8700 \kms \ corresponds to the CO sources 1, 2, 
and 3 in the giant H~II complex west of the nucleus, and to 
CO source 9 in the north of the ring.  
 The peak CO flux in these features is 80 -- 150 mJy, 
which corresponds to a beam averaged 
brightness temperature $T_{\rm b}\sim 1$\,K. 
In the interferometer map, the sources in the ring
account for 85\% of the CO flux seen with 30\,m telescope. 

The 30\,m spectrum of northwest part of ring (Fig.~1) peaks at 8250 \kms ,  
and arises in CO sources 12 and 13 on the interferometer map (Fig.~3).
The flux detected by the interferometer from these two CO sources 
is 65\% of the flux in the 30\,m beam
at the northwest ring. This may be due to these sources being  
near the half power point of the primary 
beam of the interferometer, for which no correction was made.
The high velocity tail in the 30\,m spectrum at the northwest part of the ring 
(Fig.~1) is  probably emission from NGC 1144 in the skirts of the beam. 

The interferometer maps show the CO is neither at the Seyfert nucleus, 
nor in the central disk of NGC~1144,
but rather in the north, south, and northwest parts of the ring, 
and the giant HII complex west of the nucleus (Figs.~2 and 3).
The main CO sources are
at projected radii of 3 to 6\,kpc (5$''$ to 10$''$) from the nucleus.
Except at the nucleus, 
the velocity and spatial distributions of the CO (Fig.~2) both 
agree with those of the ionized gas (H89).

Table~2 lists the CO sources on the interferometer channel maps (Fig.~2).
Most of the CO luminosity comes from 
the strong CO sources 5 to 8, in the south part of the ring,
at $v=+260$ and $+400$ \kms , relative to $cz_{\rm lsr}$ = 8750 \kms 
(the heliocentric velocity is higher by 12 \kms).
 Other prominent molecular features in the channel maps 
at $-$160 and $-$20\,\kms\ are the
CO sources 1, 2, and 3 (Fig. 3) in the giant HII region west of the nucleus. 
 These CO sources are embedded in the 3\,kpc total
extent of the giant HII complex, which
dominates the 10\,$\mu$m and H$\alpha$ emission (JG88; H89).  
It has an H$\alpha$ \ luminosity of \ $7 \times 10^{40}$ erg s$^{-1}$, five
times more than the nucleus and 10 times brighter than 
30 Doradus.  
There is no CO peak at the strong, extended radio    source 
$5''$ (3 kpc) northeast of the nucleus, which is also weak in H$\alpha$.
The radio continuum may come from supernova remnants in a 
region with low molecular gas density.

\section{NGC 1144 AS A RING GALAXY}
NGC 1144 is an excellent example of a ring galaxy,
with an elongated ring drawn out over a 20\,kpc diameter toward the 
elliptical NGC 1143.   
 Ring galaxies result from low-speed, collisions of 
compact galaxies with gas-rich disks (Lynds \& Toomre 1976;
Theys \& Spiegel 1976; Toomre 1978).    The impact
parameter must be within 15\% of the target's outer radius.
The passage is roughly along the minor axis of the disk, but 
need not be pole-on;
impacts tilted as much as 45$^\circ$ from pole-on yield rings.   
An intruder approaching in the sense of the target's spin generates 
a well-defined ring.  An approach in the retrograde sense yields a 
messy ring (Lynds \& Toomre 1976).

The Arp~118 system differs from other ring galaxies in several respects. 
 The most spectacular difference is the 1100\,\kms\ velocity range, which 
is 2 to 3 times higher than in all other known ring galaxies. 
Second,  its CO emission is unusually strong.  
 Most ring galaxies have CO luminosities ten times lower than Arp~118
(Horellou \etal 1995).  Third, its ring is a splattered mess 
with many overlapping shock waves, rather than 
a smooth, sharp ellipse like  the galaxy II~Hz~4 (Lynds \& Toomre 1976)
or the Cartwheel galaxy (e.g. Fosbury \& Hawarden 1977; Higdon 1995, 1996). 
  This suggests the
elliptical passed in the retrograde sense relative to the spiral's spin.
Fourth,  more than in most rings,  the nucleus of NGC 1144 is well off-center,
with a projected radius of only 3\,kpc to the southern and eastern CO sources,
but with radii $\ga $15 kpc to the  CO and H$\alpha$ sources in the northwest 
part of the ring.   
This implies the elliptical has a significant fraction of the
system's mass and has pushed the nucleus of the spiral 
off center.  
A good analogy to the position of the nucleus and the commotion in the 
Arp~118 ring is the retrograde model in the middle of Fig.~6 
of Lynds \& Toomre (1976). While other interpretations are possible, including a
model involving the interaction of 2 or more spirals now observed as parts of NGC1144, the
high velocities can best be reproduced  in a ``ring'' galaxy encounter.

\section{INTERPRETATION OF THE KINEMATICS}
To account for the high velocities and the H$\alpha$ ring morphology 
Hippelein (1989) made an 3-body model of a nearly face-on 
collision between a spiral and an elliptical.  
His data show 
an apparent sine-wave pattern in the H$\alpha$ velocity vs. p.a. (position
angle) in the ring, with an amplitude of 510\,\kms , 
which is confirmed by our CO observations.
By assuming the ring was circular, and inclined to us by 50$^\circ$
along a major axis at p.a. 127$^\circ$, Hippelein derived 
a rotational velocity of 670 \kms , which he noted 
would imply an unrealistically high 
dynamical mass $> 10^{12} \ {\rm M}_{\odot}$.
We suggest here an alternative to Hippelein's model that differs from his 
in the time scale and repartition of rotational and radial motions.  

In our interpretation, the disk galaxy was originally a massive, 
gas-rich spiral. 
Scaling the stellar mass derived from the $K$-band flux (JG89) to our adopted 
distance, we estimate a  mass, in \Msun , of
\begin{equation}
M(<R) = 3 \times 10^{10} \ R_{\rm kpc} \ \ \ .
\end{equation}
The $K$-band flux indicates the elliptical has half this mass (JG89).
The rotation velocity corresponding to the spiral's mass 
is 350\,\kms .  If, as in Hippelein's model,
the elliptical traversed the disk at a radius of 2\,kpc, it 
acquired a velocity of $V_{\rm esc} = \sqrt 2 V_{\rm rot}$, or 490\,\kms .
Relative to NGC 1144, the elliptical now moves toward us with its
velocity vector at 30$^\circ$ to $45^\circ$ to our line of sight, so we see a 
a line of sight component of $\sim$340\,\kms .
The elliptical is at a projected distance of 22\,kpc from the 
disk galaxy, with a true distance of 29\,kpc.
The center of mass of the pair of galaxies is two-thirds the way
from the elliptical to the spiral, or 19\,kpc, so 
closest approach occured 40\,Myr ago.

The encounter has drawn out some of the stars and gas in the former
disk of NGC 1144 in an ellipse ($R$-band image in Fig.~1).
The northwest part of the ring approaches us, the part of the ring
south of the nucleus recedes.
Since the elliptical heads at us, we look down the long
axis of the ellipse.   In ring galaxies, the encounter provokes a density wave 
that rapidly rebounds outward 
at half the intruder's velocity, i.e., at $V_{\rm rot}/\sqrt 2$,
or 250\,\kms \ for Arp 118.  This is
consistent with the ring extending half the distance to the elliptical.
The ring expansion and rotation yields an apparent line of nodes at 
p.a. 130$^\circ$ east of north on the sky (H89).  As in other ring 
galaxies, however, the minor axis must be roughly in the direction from the
spiral to the elliptical, that is, east-west.   Hence
the original disk must have been nearly edge-on to our line of sight,
with the true line of nodes nearly north-south.

We have compared the observed CO and $H\alpha$ kinematics with a model combining
rotation and expansion. We get a reasonable 
fit with a  disk kinematic major axis at p.a. $-10^\circ$
and the original disk plane inclined 20$^\circ$ to our line of sight.
The north side approaches, the south recedes.  The near side is west,
the far side is east.
The velocity shifts
north and south are the rotation component, 350\,\kms .  
The molecular clouds have the 250\,\kms
 expansion component superposed on their former rotational component
of 350\,\kms .   Because of the nearly edge-on inclination of the
former disk galaxy and the
direction of the motion of the intruder 
(probably within 45$^\circ$ of our line of sight), we see the maximum combined
effect of the expansion and rotational components of the elliptical-shaped
ring that is pointing nearly at us.
For example, CO cloud 12
 was originally in pure rotation at 300 to 350\,\kms .  
After the passage of NGC 1143, it was drawn out
into the farthest part of the ring,
following the intruder at half its speed.  Since the intruder is moving
straight toward us, the cloud's streaming component of $-$250\,\kms\   
adds vectorially
to the former motion of $-$350\,\kms \, yielding about $-$510\,\kms.
 This simple model 
 can  approximately account for the extreme linewidth by the vector
addition of the rotation, expansion and a turbulent component of about 100--150
\,\kms. 
 
In reality, the former smoothly rotating disk of NGC 1144 no longer exists
and the orbits of the gas near the nucleus 
are probably highly eccentric, more like bar streaming
than circular rotation. The molecular clouds near  the nucleus 
are moving along elliptical streamlines, with 
velocities of closest approach much higher than the circular velocity,
 of the order of  $\sqrt 2V_{\rm rot}$, or 490\,\kms .  This value is close to the red-shifted 
velocities of CO clouds 7 and 8, south of the nucleus.

\section {GAS MASS AND IR LUMINOSITY}
NGC 1144 has a very high CO luminosity, mainly as a result of 
its large velocity spread.  
We think the ratio of  H$_2$ mass to CO luminosity throughout the ring
is lower than in the Milky Way molecular clouds because the extreme velocity 
dispersion makes the CO over-luminous. 
For a mean ring radius of 6\,kpc, eq.(1) would imply an enclosed dynamical mass
of $\sim 2\times 10^{11}$\,\Msun .  If 
the gas mass is 10\% of the dynamical mass, or 
$\sim 2\times 10^{10}$\,\Msun ,  then the H$_2$ mass to CO luminosity
ratio would be two to three times lower than in  Milky Way molecular clouds.
The global $L_{\rm IR}/M$(H$_2$) ratio would be
about 10\,\Lsun /\Msun , a modest rate of star formation, but obviously 
unusual because of its simultaneous activity over a very large ring.

  Arp~118 probably differs 
from other ring galaxies in being more gas-rich and massive, 
and hence having a faster expansion velocity than other rings.  
While its ring diameter
is close to the median size for the 26 ring galaxies in the sample
by Appleton and Struck-Marcel (1987), its higher expansion velocity 
means that it is younger than most of the other known ring galaxies.
Its snowplow and shock effects will also be stronger, for the same reason,
which may explain why its star formation rate and consequent luminosity is nearly an order of
magnitude higher than for other rings.  
The encounter 
with the elliptical  
has produced a series of shock waves and an immense assembly 
of giant molecular cloud complexes and
starburst regions in a ring in the former massive, gas-rich spiral.   
Eventually the shock strength will diminish
and the massive stars will disperse the molecular clouds.  But for
the time being,  
we are apparently seeing Arp 118 near the peak of its
star forming activity.


We thank the IRAM staff at  Plateau de Bure and Pico Veleta for help with
the observations,  and J.\ Condon for his radio continuum maps. 

\clearpage
\begin{deluxetable}{lccc}
\tablecaption{Molecular Gas in Arp 118: 30m Telescope \label{tbl-1}}
\tablehead{
\colhead{Parameter}         &   \colhead{NGC 1144}      &
\colhead{NW ring}       &   \colhead{NGC 1143}}

\startdata

$\alpha_{1950}$         & $02^{\rm h}52^{\rm m}38^{\rm s}.4$ 
                        &$02^{\rm h}52^{\rm m}37^{\rm s}.3$ 
                        & $02^{\rm h}52^{\rm m}36^{\rm s}.2$ \nl
$\delta_{1950}$         & $-00^{\circ}23^{\prime}09^{\prime\prime}$ 
                        & $-00^{\circ}22^{\prime}55^{\prime\prime}$ 
                        & $-00^{\circ}22^{\prime}47^{\prime\prime}$ \nl
%
mean $cz_{\rm CO}$ (\kms ) 
                        & 8810  & 8507          &          \nl
median $cz_{\rm CO}$ (\kms) 
                        & 8750                  & --- & --- \nl
$\Delta V_{\rm CO}$(FWHM) (\kms ) 
                        & 657   & $\sim 450$    & ---   \nl
$\Delta V_{\rm CO}$(FWZI) (\kms ) 
                        & 1120  & 750 -- 1100   & ---   \nl
$I_{\rm CO}$ (K \kms )\tablenotemark{a} 
                        & 36.0  & 10.2          & $\leq1.0$ (2.5$\sigma$)\nl
$L_{\rm CO}$ (K \kms \,pc$^2$) 
                        & $8\times 10^9$        & $2\times 10^9$ 
                        & $\leq 2\times 10^8$   \nl

$I_{\rm HCN}$ (K km s$^{-1})$\tablenotemark{a} 
                        & $0.7 \pm 0.2$         & --- & --- \nl
$L_{\rm IR}$ (\ls)      & $3\times 10^{11}$     & --- & --- \nl
$L_{\rm B}$ (\ls)       & $5\times 10^{10}$     & --- &$9\times 10^{10}$  \nl

$D$ (Mpc)               & 118                   & --- & --- \nl
\enddata
\tablenotetext{a}{$T_{\rm mb}$ scale for IRAM 30m telescope; 
 at 3\,mm, $S/T_{\rm mb}=4.7$ Jy K$^{-1}$ for a point source. }
\end{deluxetable}
\clearpage


\begin{deluxetable}{lcccc}
\tablewidth{28pc}
\tablecaption{CO Sources in NGC~1144 (Arp~118): IRAM Interferometer }
\tablehead{
\colhead{CO Source}     
        &\colhead{$\Delta \alpha , \delta$ \tablenotemark{a}}   
        &\colhead{V \tablenotemark{b}} 
        &\colhead{$S_{\rm CO}$ \tablenotemark{c}}     
        &\colhead{Radius \tablenotemark{d}}     
        \nl 
\colhead{and location}
        &\colhead{(arcsec)}     
        &\colhead{(\kms )}
        &\colhead{(mJy)}
        &\colhead{(kpc)}
}
\startdata

1 west & $-$6 , $-$1 & 20 & 90       & 3.5 \nl
2 west & $-$8 , $-$3 & $-$80 & 90   & 4.6  \nl
3 west & $-$10 , $-$2 & $-$180 & 80   & 5.6  \nl
4 south & $-$7 , $-$7 & 200 ? & 50   & 7.3 \nl
5 south & $-$2 , $-$8 & 300 & 120    & 4.7  \nl
6 south & 2 , $-$5 & 380 & 170     & 3.0  \nl
7 south & 5 , $-$2 & 420 & 120   & 3.1  \nl
8 south & 6 , $-$1 & 460 & 80        & 3.5 \nl
9 north & 2 , 8 & $-$120 & 120        & 4.0  \nl
10 north& $-$1 , 10 & $-$150 & 70 ?     & 5.8  \nl
11 north & 4  ,10 & 360 & 75            & 6.7   \nl
12 NW & $-$10  , 18 & $-$510 & 80      & 11.2  \nl
13 NW & $-$19 , 19 & $-$510 & 50    & 14.5  \nl

\enddata

\tablenotetext{a}{Relative to the nucleus at
        $02^{\rm h}52^{\rm m}38^{\rm s}.7$, 
 $-00^{\circ}23'07.8''$ (1950) = $02^{\rm h}55^{\rm m}12.^{\rm s}.2$, 
 $-00^{\circ}11'00.6''$ (2000). } 
\tablenotetext{b}{Relative to $cz_{\rm lsr} =$ 8750\,\kms .} 
\tablenotetext{c}{Peak flux in individual channel maps of width 40 \kms .}
\tablenotetext{d}{Projected distance from the nucleus.}

\end{deluxetable}

\newpage

\figcaption{(Plate~1). 
{\it Left panel:} 
Images of Arp 118 in R band (top), H$\alpha + [{\rm NII}]$ 
after continuum subtraction 
(middle, Hippelein 1989), and in 1.49 GHz radio continuum (bottom, 
Condon \etal 1990) on the same scale.  
{\it Right panel:}
The CO(1--0) spectra from the IRAM 30\,m telescope at the 
three positions marked on the H$\alpha$ image by  circles indicating
the $23''$ beam.  Note the wide CO lines in NGC 1144 and the northwest part of 
the ring. The CO(2--1) spectrum 
of NGC 1143 (dashed line) shows possible weak emission centered on 8350 \kms.
Coordinates are (1950). 
\label{fig1}}

\figcaption{ Interferometer channel maps of CO(1-0) from Arp 118 at a 
resolution of 140\,\kms.  The contour step is 14\,mJy beam$^{-1}$ 
(dashed contours are negative; zero contour omitted); 
beam = $5.''3 \times 2.''5$, with $T_b/S = 6.7$\,K/Jy.  
Velocities are relative to 
 $cz_{\rm lsr} =$ 8750\,\kms. 
The cross marks the nucleus at $02^{\rm h}55^{\rm m}12.2^{\rm s}.7$, 
$-00^{\circ}11'00.6''$ (2000). 
Tick marks are $0.^{\rm s}5$ in R.A., and 10$''$ in Dec. 
\label{fig2}} 

\figcaption{(Plate~2).  Interferometer map of 
integrated CO(1--0) contours superposed  
on an H${\alpha}$ + [NII] image of Arp 118 (Hippelein 1989).  EGHR = extragalactic giant HII
region.  Coordintes are (1950). The CO map is integrated over 1285 \kms \ centered on
$cz_{\rm lsr}$  =8750\,\kms. 
The contour step is 3.9 Jy \kms \ beam$^{-1}$, with $T_b/S = 7.5$\,K/Jy. 
The cross indicates the position of the Seyfert 2 nucleus. Labels indicate the
CO sources referred to in the text and in Table 2 .
The $5.3''\times 2.5''$  beam is shown at lower left. 
\label{fig3}}

\end{document}